\documentclass[aps,pre,twocolumn,superscriptaddress]{revtex4-2}
\usepackage{graphicx}               
\usepackage{amsmath, amssymb}       
\usepackage[T1]{fontenc}            
\usepackage[utf8]{inputenc}                              
\usepackage[ justification=justified,format=plain]{caption}
\usepackage[ justification=justified,format=plain]{subcaption}
\usepackage{color}
\usepackage{physics}
\usepackage[ruled,lined]{algorithm2e}

\begin{document}

\title{Bayesian Basin Tracking: Efficient Global Continuation of Multistable Dynamical Systems}

\author{Pedro Haerter}
\email{haerter@usp.br}
\affiliation{Instituto de Física, Universidade de São Paulo, São Paulo, São Paulo, 05508-090, Brazil}

\author{Alexandre Wagemakers}
\affiliation{Nonlinear Dynamics, Chaos and Complex Systems Group, Departamento de F\'{i}sica, Universidad Rey Juan Carlos, Tulip\'{a}n s/n, 28933 M\'{o}stoles, Madrid, Spain}

\author{Alvar Daza}
\affiliation{Nonlinear Dynamics, Chaos and Complex Systems Group, Departamento de F\'{i}sica, Universidad Rey Juan Carlos, Tulip\'{a}n s/n, 28933 M\'{o}stoles, Madrid, Spain}
\affiliation{Department of Physics, Harvard University, Cambridge, MA 02138, USA}

\author{Miguel A.F. Sanjuán}
\affiliation{Nonlinear Dynamics, Chaos and Complex Systems Group, Departamento de F\'{i}sica, Universidad Rey Juan Carlos, Tulip\'{a}n s/n, 28933 M\'{o}stoles, Madrid, Spain}
\affiliation{Royal Academy of Sciences of Spain, 28004 Madrid, Spain}

\begin{abstract}
Mapping the global phase space of high-dimensional multistable dynamical systems is computationally prohibitive because conventional approaches require extensive numerical integration. Here, we introduce Bayesian Basin Tracking (BBT), an adaptive Bayesian framework that exploits the persistence of basin boundaries under parameter continuation to reconstruct global phase-space structure using only a fraction of the simulations required by conventional methods. By modeling the probability that a sampled initial condition converges to a particular attractor, the method represents the phase-space geometry established at a given parameter value through a Dirichlet–multinomial model. At a nearby parameter value, these probabilities are estimated by updating the prior distribution with newly sampled data. To detect boundary crises and bifurcations autonomously, we use the log Bayes factor as an information-theoretic sensor that triggers dense resampling only when structural changes render the historical prior statistically implausible. We validate the framework using the discrete Hénon map, the continuous-time Duffing oscillator, and a $300$-dimensional network of coupled Rössler oscillators. BBT overcomes the restrictive dimensional scaling of deterministic grid tessellations by concentrating the most computationally demanding calculations in structurally volatile regions. In high-dimensional synchronization landscapes, it achieves an almost sixfold computational speed-up while retaining theoretically derived error bounds.
\end{abstract}

\date{\today}

\maketitle

\section{Introduction}

Predicting the asymptotic state of a complex system under different parameters is a fundamental challenge across physics and engineering~\cite{pisarchik2022multistability, daza2024multistability}. Whether forecasting weather patterns, stabilizing power grids, or synchronizing coupled oscillators, these systems exhibit multistability where distinct initial conditions lead to entirely different final states. The set of initial conditions leading to a specific attractor constitutes its basin of attraction; however, determining how these basins transform over parameter changes remains a formidable computational hurdle.

Traditionally, basins of attraction are calculated by evolving a fine grid of initial conditions and comparing their asymptotic states for a specific set of parameters. Then the main features of the basins can be studied and classified for that specific regime, but as the control parameter is varied, sudden structural changes can happen within the basins. Detecting these bifurcations and analysing their effects on the principal features of the basins constitute a major challenge.

Recalculating all the points in the grid for the basins at different parameters entails a substantial workload. The computational cost increases with the dimension of the system in the form of $\varepsilon^{-d}$, where $\varepsilon<1$ is the size of the fine grid and $d$ the dimension of the phase space. While this can be affordable for simple low-dimensional maps, when we deal with high-dimensional continuous flows, the cost of numerically integrating multiple differential equations becomes computationally intractable.

Statistical sampling provides a way to avoid these problems. With Monte Carlo methods, the phase space is randomly probed and the basins are calculated within a certain precision. This procedure can reduce the computational cost, but relies on the number of points of the sample and the way they are distributed in phase space. Also, these kind of techniques can be unreliable when dealing with high-dimensional phase spaces.

Here, we introduce a new way to calculate the basins by combining both approaches, using the Bayesian framework and a sparse random sampling. We can update our knowledge and use the changes in distributions to find the characteristics of the basins as the control parameter evolves. A fundamental consideration of this work is that, in dynamical systems, basin boundaries generally deform continuously under sufficiently small parameter variations, except at bifurcations or global crises~\cite{grebogi1983crises}. Consequently, under a small parameter perturbation, it is reasonable to assume a smooth variation of the global properties, such as the relative volume of the basins and their overall structure. Using this observation, we can change the parameter slightly and compare the statistical distributions associated with the basins.

In the first step, our method constructs a probabilistic memory of the basins using a Dirichlet-Multinomial model, initialized by a single dense baseline sample. Then, as the parameter evolves, we draw only a sparse sample and compute the log-Bayes factor $\eta$  to test whether the new data are consistent with the historical distribution  or if a uniform `reset' prior is mathematically more likely. By doing so, we bypass the need for continuous dense sampling.

In this context, $\eta$ can act as an autonomous sensor. Independent of user-defined thresholds, it natively detects bifurcations and boundary crises by measuring the divergence between the historical phase-space distribution and the newly sampled data. With this new approach, it is possible to use the well-established machinery of Bayesian statistics to find some fundamental properties of the basins that are of interest, such as the relative basin volume and the basin entropy as the parameter is varied. A key advantage of the Bayesian framework lies in the rich body of work on estimation, allowing us to place our method on firm statistical grounds.

In order to illustrate the functioning of the proposed method, we apply it to classic examples from the dynamical systems literature. We reproduce some well-known results using the Bayesian Basin Tracking (BBT) approach, offering a solid starting point for researchers working in dynamical systems. Their different features also help us understand the performance of the algorithm in various scenarios. 

The present work is organized as follows. Section \ref{sec:Paradigm} reviews the current paradigm of basin continuation. Section \ref{sec:Bayes} develops the Bayesian tracking methodology. Section \ref{sec:error_analysis} provides a rigorous analysis of the error bounds. Section \ref{sec:Systems} applies the framework to classical continuous and discrete dynamical systems, as well as a high-dimensional network of coupled oscillators. Finally, Section \ref{sec:Concl} presents our conclusions.

\section{The current paradigm}
\label{sec:Paradigm}

In a dynamical system, an attractor is the final state where trajectories eventually settle. The basin of attraction is the set of all initial conditions that lead to that specific attractor~\cite{alligood_chaos_1996}. More generally, we can define a basin as the set of all initial conditions that share a common asymptotic state. For example, we can define escape basins for open Hamiltonian systems~\cite{aguirre2001wada} or basins of synchronization for networks of coupled oscillators~\cite{menck_how_2013}. 

For a dynamical system with state space $\Omega$ and evolution operator $\Phi^t$, the basin $\mathrm{B}(A)$ of an asymptotic state $A \subset \Omega$ is defined as

\begin{equation}
\mathrm{B}(A)={u_0\in\Omega\mid \lim_{t\to\infty}\mathrm{dist}(\Phi^t(u_0),A)=0},
\end{equation}
where $\Phi^t$ denotes the flow when time is continuous ($t\in\mathbb{R}_{\ge0}$) and the $t$-th iterate of a map when time is discrete ($t\in\mathbb{N}$). The boundaries of these basins can be smooth or highly complex and fractal~\cite{aguirre2009fractal}. Understanding the transformation of the basins as parameters vary provides important information about the global stability of dynamical systems. Currently, researchers use three main numerical approaches to map these basins as a parameter changes, but each faces major limitations.

The first category includes standard local continuation software, such as AUTO~\cite{doedel1981auto}, MATCONT~\cite{dhooge2003matcont}, CoCo~\cite{Dankowicz2013}, and BifurcationKit.jl~\cite{veltz:hal-02902346}. These tools are highly efficient at tracking equilibrium points and periodic orbits as a parameter varies. However, they rely strictly on linear stability. Therefore, they only analyze what happens infinitesimally close to the attractor, so they are completely blind to the global shape of the phase space and cannot detect sudden boundary changes far away from the attractor.

The second category tries to map the global space using deterministic grid searches. Programs like GAIO divide the space into small boxes to find invariant sets. While extensions of these methods~\cite{gerlach2020set} allow for parameter continuation, they suffer heavily from the ``curse of dimensionality.'' Their computing time grows as $\varepsilon^{-d}$, where $\varepsilon<1$ is the grid resolution and $d$ is the dimension of the space. This makes grid methods difficult for systems with many dimensions (i.e., $d\ge 5$). They also focus more on tracking the attractor itself rather than measuring the volume of the surrounding basin.

Aiming to tackle the dimension problem, a third category uses statistics and data. Some methods cluster time-series data to classify regions of the phase space without grids, such as bStab~\cite{Stender2021} or distance-matrix grouping~\cite{Gelbrecht2020}. However, comparing every trajectory to every other trajectory to build an $N \times N$ distance matrix is too slow for large sets of data. Nevertheless, the method in~\cite{Gelbrecht2020} solves two problems at once by grouping and identifying attractors at different parameters and measuring the phase space geometry. 

Alternatively, Monte Carlo sampling efficiently estimates global metrics, such as basin stability~\cite{Schultz2017} and basin entropy~\cite{daza2016,puy2021test}, by randomly picking points in the phase space at each parameter. However the grouping and matching of attractors are omitted from these approaches. Despite working well in high dimensions, these statistical methods fail to exploit the information gathered during previous computations. Standard Monte Carlo techniques treat the phase space at parameter $\lambda$ and $\lambda + \delta$ as completely separate problems. When the parameter changes, these methods forget everything they just learned and force the computer to resample the entire phase space from scratch, even if the basin boundaries barely moved. To fix this waste of computing resources, we need a method that uses sparse random sampling but keeps a mathematical record of the basin boundaries as they evolve.

\section{Bayesian Basin Tracking}
\label{sec:Bayes}

In this section, we present the Bayesian Basin Tracking (BBT) method, how it is constructed and how basin characteristics can be estimated within the Bayesian framework. The idea is to combine existing techniques to create a completely new algorithm that optimizes the computation of the basins for a range of parameters. Our approach also provides ways to calculate important information about the basins such as the basin stability and the basin entropy, allowing them to be classified efficiently~\cite{daza2022classifying}.


The entire method hinges on a simple assumption: the basins change smoothly for small parameter changes, except for specific critical events. These events are usually bifurcations or boundary crises~\cite{wagemakers2023}. This hypothesis allows us to track changes at parameter $\lambda + \delta$ given the information obtained at parameter $\lambda$. 

\subsection{The Bayesian probability estimator}

To construct the method, we start building a reliable oracle function $O(u)$ that labels the initial conditions $u$ according to their asymptotic state. Since evaluating the oracle requires computationally expensive numerical calculation of the trajectories, our primary objective is to mathematically minimize the number of required oracle calls.

The final state need not be an attractor; for example, a specific synchronized state can also be targeted. As labels are collected, we construct a probabilistic model for the sampling outcome. A Dirichlet prior provides a natural choice, as the probabilities $p_i$ associated with each label are updated according to the observed event counts. Replacing the classical frequentist estimate with this probabilistic representation naturally casts the estimation problem in a Bayesian framework. Because the Dirichlet prior is conjugate to the categorical (multinomial) likelihood, the posterior remains Dirichlet after applying Bayes' rule, greatly simplifying the mathematical treatment.

We assume that for $\mathcal{N}$ final states at parameter $\lambda$, the probabilities $\{p_1, p_2, \dots p_\mathcal{N}\}$ follow the prior Dirichlet distribution: 
\begin{equation}
    f(p_1, \dots, p_\mathcal{N} | \boldsymbol{\alpha}) = \frac{1}{B(\boldsymbol{\alpha})} \prod_{i=1}^\mathcal{N} p_i^{\alpha_i - 1}
\end{equation}
where $\alpha_i$ are the pseudo-counts (hyperparameters) for each category. $B(\boldsymbol{\alpha})$ is the multivariate Beta function.

After updating the parameter to $\lambda+\delta$, we inspect a new set of trajectories. 

If $c_i$ trajectories terminate in final state $i$, producing the count vector  $(c_1, c_2, \dots c_\mathcal{N})$,  we can update our belief about the distribution of probabilities $p_i$. Using Bayes' rule, the posterior distribution is simply a Dirichlet distribution with updated parameters $\alpha'_i = \alpha_i + c_i$.

This core method can be implemented directly. However, we introduce a heuristic to model the non-stationarity of the probability distribution for our problem. In dynamical systems, when parameters are varied, the basin boundaries mostly move slowly. Thus, we must take into account that the underlying probabilities are evolving, so we employ a ``fading memory'' or power prior approach with a forgetting factor $\gamma \in (0, 1)$ to discount previous information. The update rule for the prior parameters becomes:
\begin{equation}\label{alpha_update}
    \alpha'_i = \gamma \alpha_i + c_i
\end{equation}
For the first step, we choose an initial non-informative prior. A typical choice is $\alpha_i=\beta=0.5$, $\forall i$. The forgetting factor $\gamma$ plays an important role and must be tuned to keep a balance between trusting new evidence and keeping the old values for too long. We analyze the implications and tradeoffs in Sec.~IV. 

We must also deal with sudden transformations of phase space, such as bifurcations, crises, and metamorphoses during which attractors appear or disappear, basins merge, and boundaries change shape~\cite{mcdonald1985structure, grebogi1986metamorphoses, alligood_chaos_1996}. To detect when the continuity hypothesis fails, we test whether the newly observed sample is statistically consistent with the prior belief using a likelihood ratio test.

Given a sparse sample of $N_s = \sum_{i=1}^\mathcal{N} c_i$ new trajectories with observed counts $\mathbf{c} = \{c_1, \dots, c_\mathcal{N}\}$, we must determine if the underlying basin structure has undergone a transformation. Rather than utilizing a frequentist goodness-of-fit test, we compare two competing hypotheses:

\begin{enumerate}
    \item  $H_{\text{hist}}$ (Stability): The system's dynamics remain consistent with the established history. The new counts $\mathbf{c}$ are assumed to be generated by the current Dirichlet prior $\boldsymbol{\alpha}^{(\text{prior})}$.
    \item $H_{\text{reset}}$ (Crisis): The system has changed fundamentally. The data $\mathbf{c}$ is better explained by an uninformative blank slate prior $\boldsymbol{\alpha}^{(\text{reset})}$, where $\alpha_i = \beta$ for all $i$.
\end{enumerate}

The relative evidence for these hypotheses is quantified by the log Bayes factor ($\eta$). This requires calculating the marginal likelihood (evidence) of the observed counts under the Dirichlet-Multinomial distribution, derived in Appendix \ref{app:Ap1}:
\begin{equation}
    P(\mathbf{c} \mid \boldsymbol{\alpha}) = \frac{\Gamma(\sum_{i=1}^\mathcal{N} \alpha_i)}{\Gamma(N_s + \sum_{i=1}^\mathcal{N} \alpha_i)} \prod_{i=1}^\mathcal{N} \frac{\Gamma(c_i + \alpha_i)}{\Gamma(\alpha_i)}
    \label{eq:marginal_likelihood}
\end{equation}

For numerical stability, we work with the log-evidence $L(\boldsymbol{\alpha}) = \ln P(\mathbf{c} \mid \boldsymbol{\alpha})$. Using the multivariate Beta function representation, $L(\boldsymbol{\alpha}) = \ln \text{B}(\boldsymbol{\alpha} + \mathbf{c}) - \ln \text{B}(\boldsymbol{\alpha})$, the decision criterion is defined as:
\begin{equation}
    \eta = L(\boldsymbol{\alpha}^{(\text{prior})}) - L(\boldsymbol{\alpha}^{(\text{reset})})
    \label{eq:log_bayes_factor}
\end{equation}

When $\eta<0$ the $H_\text{reset}$ hypothesis is better supported by the observed data and the algorithm triggers an autonomous reset: it discards the historical priors entirely and samples the region densely to re-learn the distribution from scratch. This means that the weak, uninformative reset model provides a better explanation of the data than the historical model. This transition is self-calibrating: as the accumulated evidence increases (large $\sum \alpha_i$), increasingly larger discrepancies in $\mathbf{c}$ are required to modify the posterior, whereas weaker priors adapt more readily to new data.

Once the probabilities inside a box are updated—either through the standard Bayesian update or following an autonomous reset, we can derive some characteristics of the basins, such as volume fractions or the basin entropy~\cite{daza2016}. 



\subsection{Continuation algorithm} 

We focus on a single box $\mathcal{B}$ of the state space delimited by minimum and maximum coordinates $[x^i_{min}, x^i_{max}]$ along each dimension $i$, so that the algorithm will only take into account basins in this region. We assume that the oracle $O(u)$ with $u\in \mathcal{B}$ has been set up and is available for each parameter.

The procedure, detailed in Algorithm \ref{alg:bbt_box}, begins with a dense initialization to establish a high-confidence ground truth. For subsequent parameter increments, the algorithm relies on sparse sampling and the Log-Bayes factor $\eta$ to autonomously govern the sampling density. If a new basin appears within the box, the method naturally accommodates it by assigning a new categorical label; if the sudden appearance of this final state renders the sparse counts mathematically improbable under the historical prior, the algorithm autonomously triggers $\eta<0$ and executes the dense resampling protocol to restore accuracy.

\begin{figure*}
    \includegraphics[width = \textwidth]{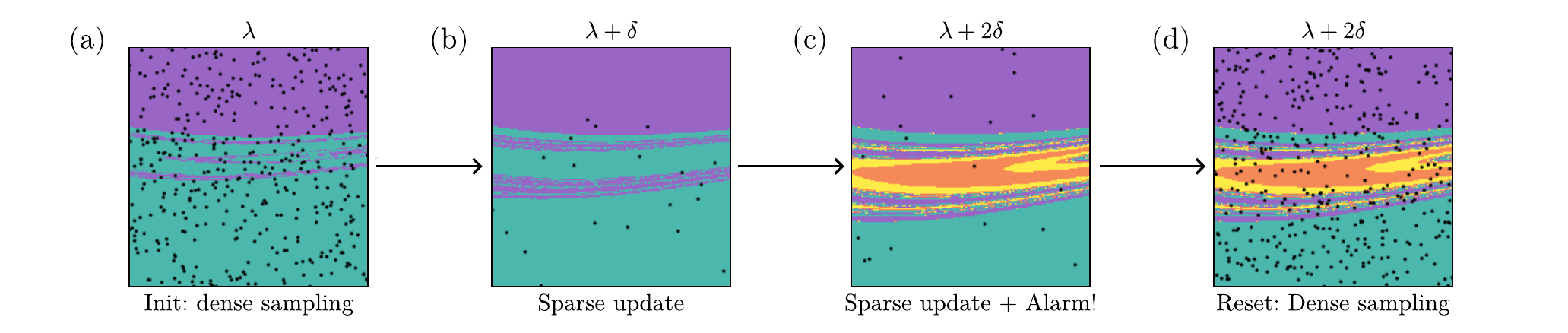}
    \caption{ Example of typical steps of the algorithm for a given box of the state space. (a) Initialization of the priors using dense sampling, represented by black dots. (b) The parameter changes to $\lambda + \delta$ and the priors are updated with a sparse sampling. (c) Change of the parameter to $\lambda + 2\delta$, but a bifurcation has occurred and the basins completely changed. The hypothesis that the sample is consistent with the historical prior is rejected. The algorithm switches to dense sampling routine in (d) to refresh the probabilities. \label{fig1}}
\end{figure*}

These steps are illustrated in Fig.~\ref{fig1} using 2D basins. The sparse routine sampling is disrupted when the statistical test rejects the hypothesis that the sample originated from the historical distribution. Using this methodology, we can calculate important features of the basins, such as the individual basin-volume fractions and the basin entropy.

\begin{algorithm}
\caption{Bayesian Basin Tracking for a single box $\mathcal{B}$}
\label{alg:bbt_box}
\SetKwInOut{Input}{input}
\Input{Box $\mathcal{B}$, dense sample size $N_d$, sparse sample size $N_s$, prior base $\beta$, forgetting factor $\gamma$}
\Begin{
    \textbf{Initialization at $\lambda_0$:}\;
    Sample $N_d$ trajectories uniformly in $\mathcal{B}$\;
    Obtain categorical counts $\mathbf{c}^{(dense)}$\;
    Initialize prior: $\boldsymbol{\alpha} \leftarrow \mathbf{c}^{(dense)} + \beta$\;
    
    \vspace{0.2cm}
    \textbf{Continuation to $\lambda + \delta$:}\;
    Track and match valid attractors from $\lambda$ to $\lambda + \delta$\;
    Attenuate historical memory: $\boldsymbol{\alpha}^{(prior)} \leftarrow \gamma \boldsymbol{\alpha}$\;
    Sample $N_s$ trajectories sparsely in $\mathcal{B}$\;
    Obtain new sparse counts $\mathbf{c}^{(sparse)}$\;
    
    \vspace{0.2cm}
    \textbf{Hypothesis Testing:}\;
    Compute $L(\boldsymbol{\alpha}^{(prior)}) \leftarrow \ln P(\mathbf{c}^{(sparse)} \mid \boldsymbol{\alpha}^{(prior)})$\;
    Compute $L(\boldsymbol{\alpha}^{(reset)}) \leftarrow \ln P(\mathbf{c}^{(sparse)} \mid \beta)$\;
    $\eta \leftarrow L(\boldsymbol{\alpha}^{(prior)}) - L(\boldsymbol{\alpha}^{(reset)})$\;
    
    \vspace{0.2cm}
    \textbf{Decision and Update:}\;
    \eIf{$\eta < 0$}{
        \tcp{Discard prior and trigger dense resampling}
        Sample $N_d$ trajectories in $\mathcal{B} \rightarrow \mathbf{c}^{(dense)}$\;
        $\boldsymbol{\alpha} \leftarrow \mathbf{c}^{(dense)} + \beta$\;
    }{
        \tcp{Update posterior via Bayes' rule}
        $\boldsymbol{\alpha} \leftarrow \boldsymbol{\alpha}^{(prior)} + \mathbf{c}^{(sparse)}$\;
    }
    
    \vspace{0.2cm}
    Calculate desired quantifiers such as the expected local entropy $\mathbb{E}[S]$ and its variance $\text{Var}(S)$ using $\boldsymbol{\alpha}$\;
}
\end{algorithm}

\subsection{Computing the basin stability and the basin entropy}

We evaluate the fraction of phase space volume $V$ occupied by each basin. The expected volume of basin $i$ in a specific box is $\tilde{V}_i = V_{\text{box}} \cdot \mathbb{E}[p_i]$, where the expected probability from the Dirichlet posterior $\mathbb{E}[p_i]$ can be calculated as the mean of the distribution:
\begin{equation}\label{exp_p_i}
    \mathbb{E}[p_i] = \frac{\alpha'_i}{\sum_j \alpha'_j} = \frac{\gamma \alpha^{(prior)}_i + c_i}{\sum_j \alpha'_j}.
\end{equation}
By adding these volumes across all $N_b$ boxes, we obtain the global basin stability (volume fraction) for the entire phase space.


Beyond basin volumes, the basin entropy provides another useful measure of the uncertainty of the basins~\cite{daza2016, daza2022classifying}. One of the key advantages of using a Bayesian approach is the existence of closed-form expressions for the moments of the Shannon entropy. The expected entropy $\mathbb{E}[S]$ of the posterior distribution is given by~\cite{wolpert1995,archer2014}:
\begin{equation}
    \mathbb{E}[S] = \psi(\alpha_0 + 1) - \sum_{i=1}^\mathcal{N} \frac{\alpha'_i}{\alpha_0} \psi(\alpha'_i + 1),
    \label{eq:bayes_entropy}
\end{equation}
where $\alpha_0 = \sum_{i=1}^\mathcal{N} \alpha'_i$ is the total count mass, and $\psi(\cdot)$ is the digamma function. This serves as a robust replacement for the ``plug-in'' estimator $H = -\sum \hat{p}_i \log \hat{p}_i$. Both estimators of the entropy carry negative bias at sparse sample sizes. 

Then the basin entropy is defined as the spatial average of the Shannon entropy in Eq.~(\ref{eq:bayes_entropy}) over a tessellation of phase space. We assume a covering of $N_b$ non-overlapping boxes. Since the boxes are statistically independent (given the parameters), the total basin entropy $S_b$ for parameter $a$ can be determined as the average of the local entropies:
\begin{equation}
    S_b^a = \frac{1}{N_b}\sum_{k=1}^{N_b} \mathbb{E}[S_k],
\end{equation}
and its variance can be derived by summing the variances of the independent boxes:
\begin{equation}
    \text{Var}(S_b^a) = \frac{1}{N_b^2} \sum_{k=1}^{N_b} \text{Var}(S_k).
\end{equation}

One of the main advantages of the Bayesian approach is the availability of a closed formula for the variance of the entropy~\cite{archer2014}, which helps us quantify the uncertainty of our estimate without further sampling. The formula for the variance and its scaling behavior are studied in the next section. 


\section{Estimator analysis and error bounds}
\label{sec:error_analysis}

To ensure the reliability of the global continuation method, it is crucial to establish theoretical bounds on the estimation errors. In this section, we analyze the statistical error of the Bayesian estimators as a function of the number of samples ($N_d$, $N_s$), the number of final states of the system ($\mathcal{N}$), and the dimension of the phase space ($d$).

\subsection{Error bounds on basin volume fractions}

In the Dirichlet-multinomial model, the probability $p_i$ of a trajectory reaching final state $i$ within a specific box is treated as a random variable. The quality of our estimate depends on the variance of this variable.

During the \textit{initialization} step at a parameter $\lambda_0$, we sample the box densely with $N_d$ points. Assuming a symmetric non-informative prior base $\beta$ (e.g., $\beta = 0.5$), the posterior hyperparameters become $\alpha_i = \beta + c_i$, where $c_i$ is the count of trajectories ending in attractor $i$. The total pseudo-count mass is $\alpha_0 = \mathcal{N}\beta + N_d$. 

The posterior variance of $p_i$ is given by the expected Bayesian volume fraction $\hat{p}_i = \mathbb{E}[p_i]$ is given by:
\begin{equation}
    \text{Var}(\hat{p}_i) = \frac{\alpha_i(\alpha_0 - \alpha_i)}{\alpha_0^2 (\alpha_0 + 1)} \approx \frac{\hat{p}_i (1 - \hat{p}_i)}{N_d}.
\end{equation}
As expected, the error at the dense sampling stage scales as $\mathcal{O}(1/\sqrt{N_d})$. To quantify the uncertainty of the estimator, we derive probabilistic error bounds using concentration inequalities.

If we want to bound the error of a specific basin volume fraction, we treat the Monte Carlo evaluation as a sum of $N_d$ independent Bernoulli indicator variables $X_j \in \{0, 1\}$, where $X_j = 1$ if the $j$-th sampled trajectory falls into basin $i$, and $0$ otherwise. The empirical probability is the sample mean $\hat{p}_{i, emp} = \frac{1}{N_d} \sum X_j$. Applying Hoeffding's inequality~\cite{hoeffding1963probability} to this sum, the probability that the estimate deviates from the true probability $p_i^*$ by more than a margin $\epsilon$ is bounded by:
\begin{equation}
    P\left( |\hat{p}_{i, emp} - p_i^*| \geq \epsilon \right) \leq 2 \exp\left(-2 N_d \epsilon^2\right).
\end{equation}
Given that the Bayesian estimate $\hat{p}_i$ converges to $\hat{p}_{i, emp}$ for large $N_d$, this bound provides a criterion for determining the required sample size for the initialization step.

If we instead consider the simultaneous estimation of all $\mathcal{N}$ final states, the $L_1$ distance between our estimated categorical distribution $\mathbf{\hat{p}}$ and the true distribution $\mathbf{p}^*$ is bounded by the Bretagnolle-Huber-Carol inequality~\cite{agresti2012categorical}, leading to
\begin{equation}
    P\left( \sum_{i=1}^\mathcal{N} |\hat{p}_i - p_i^*| \geq \epsilon \right) \leq 2^\mathcal{N} \exp\left( - \frac{N_d \epsilon^2}{2} \right).
    \label{eq:bhc_bound}
\end{equation}
Equation~\eqref{eq:bhc_bound} provides a principled criterion for selecting the dense sample size $N_d$. To guarantee a total categorical error below $\epsilon$ with confidence $1 - \delta$, the required sample size scales linearly with the number of possible final states. Since the number of coexisting final states is typically modest in practical applications, the initialization cost remains computationally tractable.


\subsection{Error bounds on local and global basin entropy}

Bounding the statistical error of the Shannon entropy is notoriously difficult due to the singularity near zero of the logarithm, which causes standard plug-in estimators to be systematically biased. The Bayesian estimator employed in our method (Eq.~\ref{eq:bayes_entropy}) mitigates this bias and provides a direct quantification of the uncertainty through the exact closed-form variance derived by Wolpert and Wolf~\cite{wolpert1995} :

\begin{equation}\label{eq:var_bayes_DP}
\begin{split}
    \mathbb{E}[S^2] &= \sum_{i=1}^K \frac{\alpha_i(\alpha_i + 1)}{\alpha_0(\alpha_0 + 1)} \biggl[ \big(\psi(\alpha_i + 2) - \psi(\alpha_0 + 2)\big)^2 \\ 
    &\quad + \psi'(\alpha_i + 2) - \psi'(\alpha_0 + 2) \biggr] \\
    &\quad + \sum_{i \neq j} \frac{\alpha_i \alpha_j}{\alpha_0(\alpha_0 + 1)} \biggl[ \big(\psi(\alpha_i + 1) - \psi(\alpha_0 + 2)\big) \\ 
    &\quad \times \big(\psi(\alpha_j + 1) - \psi(\alpha_0 + 2)\big) - \psi'(\alpha_0 + 2) \biggr].
\end{split}
\end{equation}
While exact, this formulation is mathematically dense. To expose the scaling behavior, we use the multivariate Delta method. For a large total pseudo-count mass $\alpha_0 = \sum_{i=1}^\mathcal{N} \alpha_i$, the Taylor expansion of the entropy around the posterior mean $\hat{p}_i = \alpha_i/\alpha_0$ yields the local variance in box $\mathcal{M}$:
\begin{equation}
    \text{Var}(S_\mathcal{M}) \approx \frac{1}{\alpha_0}\left[\sum_{i=1}^\mathcal{N} \hat{p}_i(\ln \hat{p}_i)^2 - \left(\sum_{i=1}^\mathcal{N} \hat{p}_i \ln \hat{p}_i\right)^2\right] = \frac{\sigma_{H,\mathcal{M}}^2}{\alpha_0}.
    \label{eq:delta_variance}
\end{equation}
The numerator $\sigma_{H,\mathcal{M}}^2$ is the variance of the surprisal $-\ln p_i$ under the categorical distribution defined by $\hat{p}_i$. It is fully determined by the proportions of the basins inside the box: it vanishes when a single attractor dominates ($\hat{p}_i \to 1$ for some $i$) and is largest when multiple basins coexist at comparable probabilities.

At steady state in the smooth continuation regime, the effective number of samples can be written as $\alpha_0 = N_s/(1-\gamma)$ (see Appendix~\ref{app:logBF}), so the local variance becomes:
\begin{equation}
    \text{Var}(S_\mathcal{M}) \approx \frac{\sigma_{H,\mathcal{M}}^2\,(1-\gamma)}{N_s}.
    \label{eq:local_var_steady}
\end{equation}

As for the variance of the global basin entropy $S_b$ calculated for a tessellation of $N_b$ independent boxes, we have
\begin{equation}
    \text{Var}(S_b) = \frac{1}{N_b^2}\sum_{\mathcal{M}=1}^{N_b}\text{Var}(S_\mathcal{M}) = \frac{(1-\gamma)\,\bar{\sigma}_H^2}{N_b\, N_s},
    \label{eq:global_variance}
\end{equation}
where $\bar{\sigma}_H^2 = \frac{1}{N_b}\sum_{\mathcal{M}=1}^{N_b}\sigma_{H,\mathcal{M}}^2$ is the average surprisal variance across boxes. The global estimation error therefore scales as $\mathcal{O}(1/\sqrt{N_b\, N_s})$, modulated by $\sqrt{1-\gamma}$ and the average information content of the basin structure.

This scaling reveals a natural error budget. Only boxes intersecting basin boundaries contribute significantly to the global variance. Boxes entirely contained within a single basin have $\sigma_{H,\mathcal{M}}^2 \approx 0$ and contribute negligibly to the global variance. Only boxes located at basin boundaries, where multiple fates coexist, carry significant uncertainty. The quantity $\bar{\sigma}_H^2$ is not a free parameter; rather, it is computed from the posterior means at each step, providing a direct and adaptive estimate of the global error alongside the entropy itself.

\subsection{Balance between forgetting factor and event detection}

As previous information fades with the forgetting factor $\gamma$, a natural tension arises: if we forget too quickly, every small change looks like a crisis. We need to check that a smooth parameter change does not trigger a false reset through the log Bayes factor test. 

To verify that a reasonable drift of the probabilities does not trigger false alarms, we develop a series of approximations. All expressions simplify when written in terms of the memory ratio $\rho = \gamma/(1-\gamma)$. First, we approximate the lag in the estimated probabilities due to the forgetting factor. We then estimate the expected log Bayes factor using the mismatch between the estimation and the true probabilities through the Kullback-Leibler divergence.

If the true probabilities drift linearly, $p_i(\lambda + \delta) = p_i(\lambda) + \delta\, \partial p_i / \partial a$, the  weighted posterior mean lags behind the truth. Unrolling the steady-state recurrence $\hat{p}_i^{(n+1)} = \gamma\,\hat{p}_i^{(n)} + (1-\gamma)\,p_i(n)$ yields (see Appendix~\ref{app:logBF}):
\begin{equation}
    \hat{p}_i^{(n)} = p_i(n) - \rho\,\delta\,\frac{\partial p_i}{\partial a}.
    \label{eq:drift_lag}
\end{equation}
The mismatch between the prior prediction $\hat{\mathbf{p}}$ and the true distribution $\mathbf{p}$ is measured by the Kullback--Leibler divergence. Using the divergence and the previous lag estimation, we obtain that
\begin{equation}
    D_{KL}(\mathbf{p}\|\hat{\mathbf{p}}) \;\approx\; \frac{\rho^2\delta^2}{2}\,\sum_{i=1}^\mathcal{N} \frac{\left(\frac{\partial p_i}{\partial a}\right)^2}{\hat p_i}.
    \label{eq:dkl_drift}
\end{equation}

Now we turn our attention to the log Bayes factor. We want to estimate the expectation in the steady state drift regime. The factor $\eta$ compares two models for the newly observed counts: the history model (prior built from accumulated evidence) and the reset model (a weak uninformative prior $\alpha_i = \beta = 1/2$). In Appendix~\ref{app:logBF} we derive the leading-order approximation:
\begin{equation}
    \mathbb{E}[\eta] \;\approx\; \underbrace{\frac{\mathcal{N}-1}{2}\ln\frac{2N_s}{\mathcal{N}}}_{\text{Occam advantage}} \;-\; \underbrace{N_s\,D_{KL}(\mathbf{p}\|\hat{\mathbf{p}})\!\left(1 - \frac{1}{\rho}\right)}_{\text{drift penalty}}.
    \label{eq:eta_main}
\end{equation}
The first term accounts for the structural advantage of the history model while the second term is the cost of the tracking lag, attenuated by the factor $(1-1/\rho)$: for $\rho \gg 1$ (long memory), the full $D_{KL}$ penalty applies. As $\rho \to 1$ (rapid forgetting), the prior carries so few pseudo-counts that it accommodates the mismatch almost for free. The expansion requires $\rho > 1$ (equivalently $\gamma > 1/2$).

False resets are avoided ($\mathbb{E}[\eta] > 0$) when the drift penalty is smaller than the Occam advantage. Substituting the lag from Eq.~\eqref{eq:dkl_drift}:
\begin{equation}
    \frac{N_s\,\rho(\rho-1)\,\delta^2}{2}\, \sum_{i=1}^\mathcal{N} \frac{\left(\frac{\partial p_i}{\partial a}\right)^2}{\hat p_i} \;\ll\; \frac{\mathcal{N}-1}{2}\ln\frac{2N_s}{\mathcal{N}}.
    \label{eq:self_consistency}
\end{equation}
The left-hand side grows quadratically with the step size $\delta$ and linearly with $N_s$, while the right-hand side grows only as $\ln N_s$. The factored form $\rho(\rho-1)$ makes the validity boundary $\rho > 1$ immediately transparent. In practice, the two sides are separated by many orders of magnitude.

As a concrete check: for $\gamma = 0.7$ ($\rho \approx 2.33$), $N_s = 20$, $\mathcal{N} = 3$, and probabilities $(0.5, 0.3, 0.2)$ drifting at rates $(0.1, -0.05, -0.05)$ per unit change in the parameter with step $\delta = 0.01$, the Occam advantage is $\frac{\mathcal{N}-1}{2}\ln(2N_s/\mathcal{N}) = \ln(40/3) \approx 2.6$, while the drift penalty is $\approx 1.3 \times 10^{-4}$, four orders of magnitude smaller.

\subsection{Computational gain} 

The whole point of the method is to save computational power while keeping the metrics within a controlled error range. To quantify this gain, it is useful to define the average alarm rate probability $\bar p_a$ over a parameter sweep, which is simply the fraction of parameter steps at which an alarm is triggered. This probability provides a direct indicator of the computational gain compared to the full Monte Carlo simulation. The total number of samples required by the Bayesian framework over $T$ parameter steps and $N_b$ boxes is $N_t = T N_b (N_s (1-\bar p_a)  + N_d \bar p_a)$. Compared to a brute-force dense grid requiring $N_d N_bT$ samples, the computational gain is: 
\begin{equation}
    G = \frac{N_d N_b T}{T N_b (N_s (1-\bar p_a)  + N_d \bar p_a )} \approx \frac{N_d}{N_s + N_d \bar p_a}.
\end{equation}
The use of the method is justified as long as $\bar p_a \ll 1$. In other words, as long as the boundaries remain relatively stable for wide parameter regimes, the method will perform efficiently. The examples given in the following section show that this is naturally the case for dynamical systems, providing a significant computational improvement.

An interesting feature of the method lies in its increasing efficiency as the parameter region is explored more finely. This occurs because the parameter values where bifurcations and crises take place are usually scattered following a fractal distribution. Therefore, the probability of alarm behaves as $p_a\propto \varepsilon_p^{\alpha}$, where $\alpha_p=D_p-d_a \leq 1$ is the difference between the dimension of the parameter region being probed $D_p$ and the fractal dimension of the alarm events $d_a$, and $\varepsilon_p<1$ represents the grid used for the parameter region. The gain increases when it is most needed, that is, at finer explorations of the parameter regions. Nevertheless, other computational limitations may appear as the resolution is increased, and the gain is always bounded by the ratio $N_d/N_s$.

It is also easy to see that increasing the dense sample size $N_d$ can inflate the theoretical gain $G$, although in a somewhat artificial manner. In any case, the crucial point is that the adaptive nature of the log-Bayes factor ensures that computational resources are allocated exclusively to structurally critical parameter regions, regardless of the absolute scale of the system.


\section{Dynamical systems application}
\label{sec:Systems}

To evaluate the efficiency of the proposed method, we apply the procedure to three different dynamical systems, with complementary characteristics and different classes of complexity. First, we study the Hénon map~\cite{henon1976two}, a 2D discrete-time map exhibiting a wide variety of dynamical behaviors, including chaos and multistability~\cite{gallas1993structure}. Second, we examine the paradigmatic Duffing oscillator~\cite{kovacic2011duffing}, defined in continuous-time and presenting a wide range of nonlinear phenomena such as period-doubling and basin erosion. Finally, a network of coupled Rössler oscillators helps us illustrate how the methodology scales and adapts to the estimation of basins in high-dimensional state spaces.

\subsection{Hénon map}

Despite its simple algebraic form, the Hénon map is a paradigmatic model for the study of basins of attraction. Its dynamics range from simple fixed points and periodic orbits to complex, fractal chaotic attractors, making it an ideal benchmark for our Bayesian framework. The map equations used for this study are given by~\cite{henon1976two,alligood_chaos_1996}: 
\begin{align}
  x_{n+1} &= a - x_n^2 - b y_n, \\
  y_{n+1} &= x_n.
\end{align}

By fixing $b=-0.3$, the map is area-contracting, allowing for the emergence of different attractors. We use the BBT method to probe the parameter range $a \in [0, 2.0]$, where the system transitions from periodic behavior to fully developed chaos. Before starting the continuation algorithm, we must set up the attractor detection infrastructure and the software necessary to match the attractors from one parameter to the next. This step lies beyond the scope of the present work, but it is nevertheless essential to the method. For this particular example, we have used the Julia numerical library Attractors.jl that implements the algorithms introduced in~\cite{Datseris2022,datseris2023}.

\begin{figure}[ht!]
  \centerline{\includegraphics[width=0.50\textwidth]{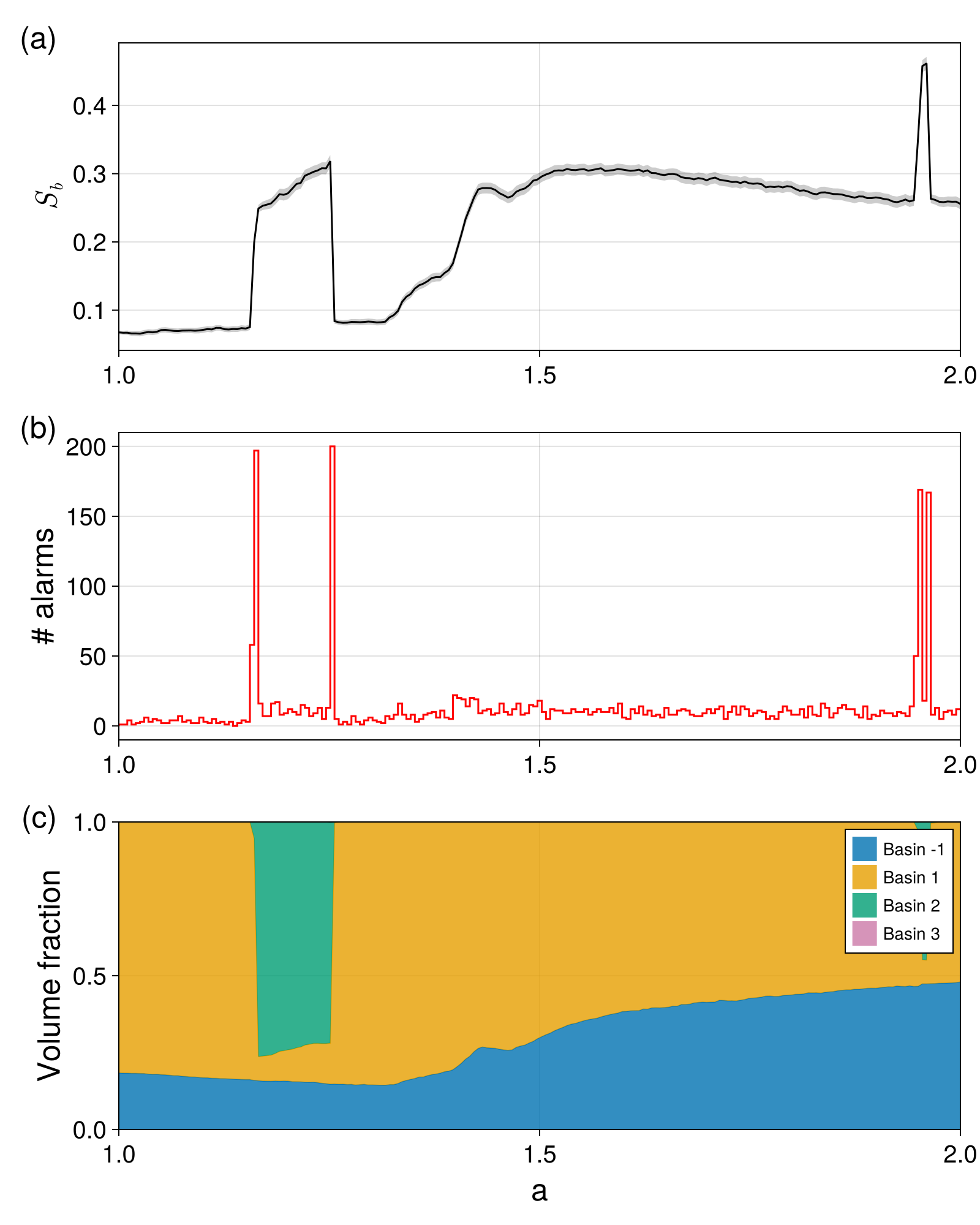}}
    \caption{Computation of basin properties as a function of the parameter $a$ for the Hénon map. (a) The black line represents the estimated basin entropy $S_b$, and the variance of the estimator is marked with a light gray band. (b) The number of alarms, i.e., the number of boxes where the dense sampling is performed, at each value of the parameter $a$. (c) The relative volume of each basin, where the label $-1$ denotes trajectories escaping to infinity. Parameters for this simulation are: $\gamma = 0.7$, $N_s = 15$, $N_d = N_s^2$, $N_b = 225$, $b = -0.3$ and 200 steps for the parameter $a$.}\label{fig:fig2}

\end{figure}

Figure~\ref{fig:fig2} summarizes the results obtained using the BBT. We display the global basin entropy $S_b$ in (a), the number of alarms per parameter step in (b) and the relative volume of each basin in panel (c). Our results are in strong agreement with published data characterizing the basins of the Hénon map~\cite{wagemakers2023}. The calculated basin sizes accurately reflect the bifurcations and crises consistently observed in this system.

The sharp peaks in Fig.~\ref{fig:fig2}(b) identify the specific parameters where significant structural changes occur, demonstrating the algorithm's capability to detect basin transitions with minimal sampling. For most of the parameter range, the number of dense-sampling events remains low. We can compute the average computational gain from this figure by estimating the alarm rate probability $\bar p_a = 0.055$, which yields a computational gain of $G = 8.15$ for this simulation.

\subsection{Duffing oscillator}

The second paradigmatic system considered is the Duffing oscillator, which represents a particle moving in a cubic potential under damping and periodic forcing. The equation that describes its motion reads

\begin{equation}
	\ddot{x}+\delta\dot{x}-x+x^3=F\sin(\omega t),
\end{equation}
where $\delta$ represents the damping coefficient, and $F$ and $\omega$ are the amplitude and frequency of the forcing, respectively. 

By fixing $\delta=0.2$ and $F=0.2$ and varying the forcing angular frequency $\omega$, the Duffing oscillator exhibits several dynamical regimes~\cite{kovacic2011duffing}. We compute the basins over the range $\omega\in[0,1.5]$, where the system transitions from periodic to chaotic dynamics, with the emergence of multiple attractors.

\begin{figure}[ht!]
  \centerline{\includegraphics[width=0.50\textwidth]{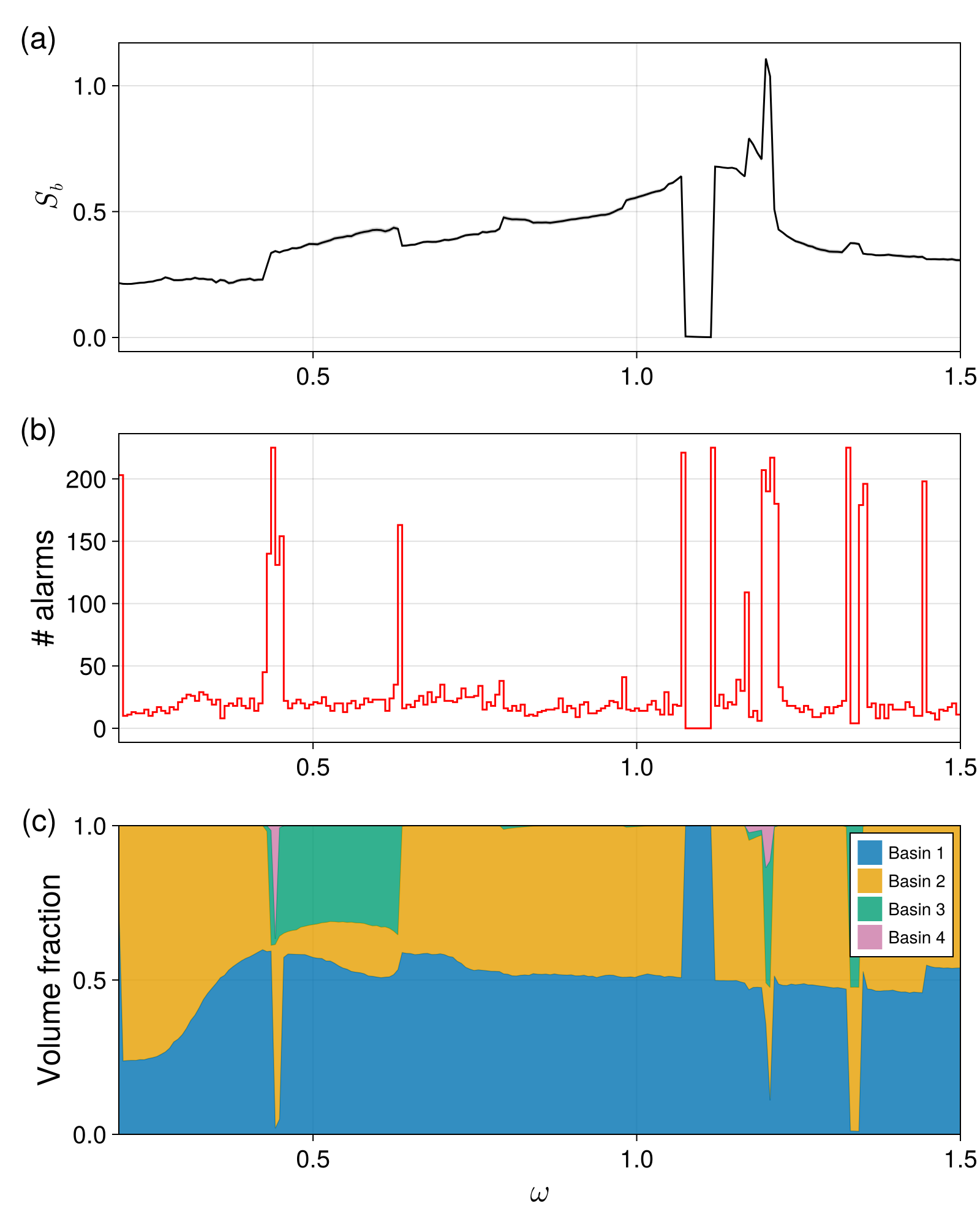}}
    \caption{Computation of the basin entropy as a function of the parameter $\omega$ for the Duffing oscillator. (a) The black line represents the estimated entropy $S_b$. Unlike for Fig.~\ref{fig:fig2}, here the variance is smaller than the thickness of the line, so it cannot be seen in this plot. (b) Number of alarms for each value of the parameter $\omega$. (c) Relative basin volumes. Parameters for this simulation are: $\gamma = 0.7$, $N_s = 15$, $N_d = N_s^2$, $N_b = 225$, $\delta = 0.2$, $F=0.2$ and 200 steps for the parameter $\omega$.\label{fig:fig3}}
\end{figure}

Figure~\ref{fig:fig3} shows the results obtained via the BBT. As for the Hénon map, we present the basin entropy, the relative volume of each basin, and the number of times that the autonomous reset was called per step. This quantity represents the structural changes occurring in each box of the partition, with a high number of calls when the system passes through transitions, such as crises or bifurcations, and values close to zero for relatively stationary regions of the parameter space. For this system and this choice of parameters, the average alarm rate is $\bar p_a = 0.14$ resulting in a computational gain $G = 5.2$.

\subsection{Coupled R\"ossler oscillators on a small-world network}

 We now consider a setting that combines high dimensionality with a non-trivial synchronization landscape: a network of R\"ossler oscillators coupled through a Watts-Strogatz small-world topology~\cite{watts1998collective}. This example is inspired by the basin stability analysis of Menck and Kurths~\cite{menck_how_2013}, who showed that the balance between network topology and coupling strength governs the synchronization of this system. 

The system under study consists of $N$ diffusively coupled R\"ossler oscillators on a connected graph with Laplacian matrix $L$:
\begin{align}
  \dot{x}_i &= -y_i - z_i - K\sum_{j=1}^{N} L_{ij}\, x_j, \nonumber\\
  \dot{y}_i &= x_i + a\, y_i, \label{eq:rossler_network}\\
  \dot{z}_i &= b + z_i(x_i - c). \nonumber
\end{align}
The standard parameters values $a=0.2$, $b=0.2$, and $c=9.0$ place each oscillator in the chaotic regime. The coupling acts through the $x$-variable only, with strength $K>0$.

The synchronous state, in which all oscillators follow the same trajectory, can be analyzed via the Master Stability Function (MSF) framework~\cite{Pecora1998}. In particular, the synchronous manifold remains linearly stable only if all transverse modes $K\lambda_j$, $j=2,\ldots,N$, fall within the MSF stability interval $(\alpha_1,\alpha_2)$, where $\lambda_j$ are the non-zero Laplacian eigenvalues. For these parameter values, $\alpha_1 \approx 0.123$ and $\alpha_2 \approx 4.663$, the authors in~\cite{Boccaletti2002} derived the condition on the coupling $K \in I_s = (\alpha_1/\lambda_2,\, \alpha_2/\lambda_N)$.

The underlying graph is generated by the Watts-Strogatz model with $N=100$ nodes and mean degree $\langle k \rangle = 8$. For a fixed rewiring probability $p$, the algorithm proceeds as follows: the graph is constructed, the Laplacian spectrum is computed, and the interval $I_s$ of admissible values of $K$ is determined. If $I_s$ is empty (the eigenvalue ratio $R$ exceeds the MSF bound), synchronization is linearly impossible, and that network realization is excluded from the continuation analysis. Otherwise, the Bayesian continuation is performed over a range of values in $K \in I_s$.

At each value of $K$, the algorithm integrates the full $3N$-dimensional ODE from a random initial condition $\mathbf{u}_0$ drawn uniformly from the bounding region ($x_i, y_i \in [-12,12]$, $z_i \in [-8,35]$). After discarding a transient, the Golomb--Rinzel synchrony measure~\cite{Golomb1993}
\begin{equation}
    r = \frac{\operatorname{Var}_t\!\left(\langle x_i(t) \rangle_i\right)}{\left\langle \operatorname{Var}_t(x_i(t)) \right\rangle_i}
\end{equation}
is computed over a measurement window, where $\langle\cdot\rangle_i$ denotes the spatial mean over oscillators and $\operatorname{Var}_t$ the temporal variance. The initial condition is classified as synchronizing if $r > 0.9$, and as desynchronizing otherwise. 

The state space is $3N=300$-dimensional, making any grid-based tessellation infeasible. We therefore use a single box covering the entire bounding region. The Bayesian continuation sweeps $K$ across $I_s$ in $50$ steps, using $N_s=20$ sparse samples per step and $N_d=200$ samples for initialization or autonomous reset.

\begin{figure}[ht!]
  \centerline{\includegraphics[width=0.50\textwidth]{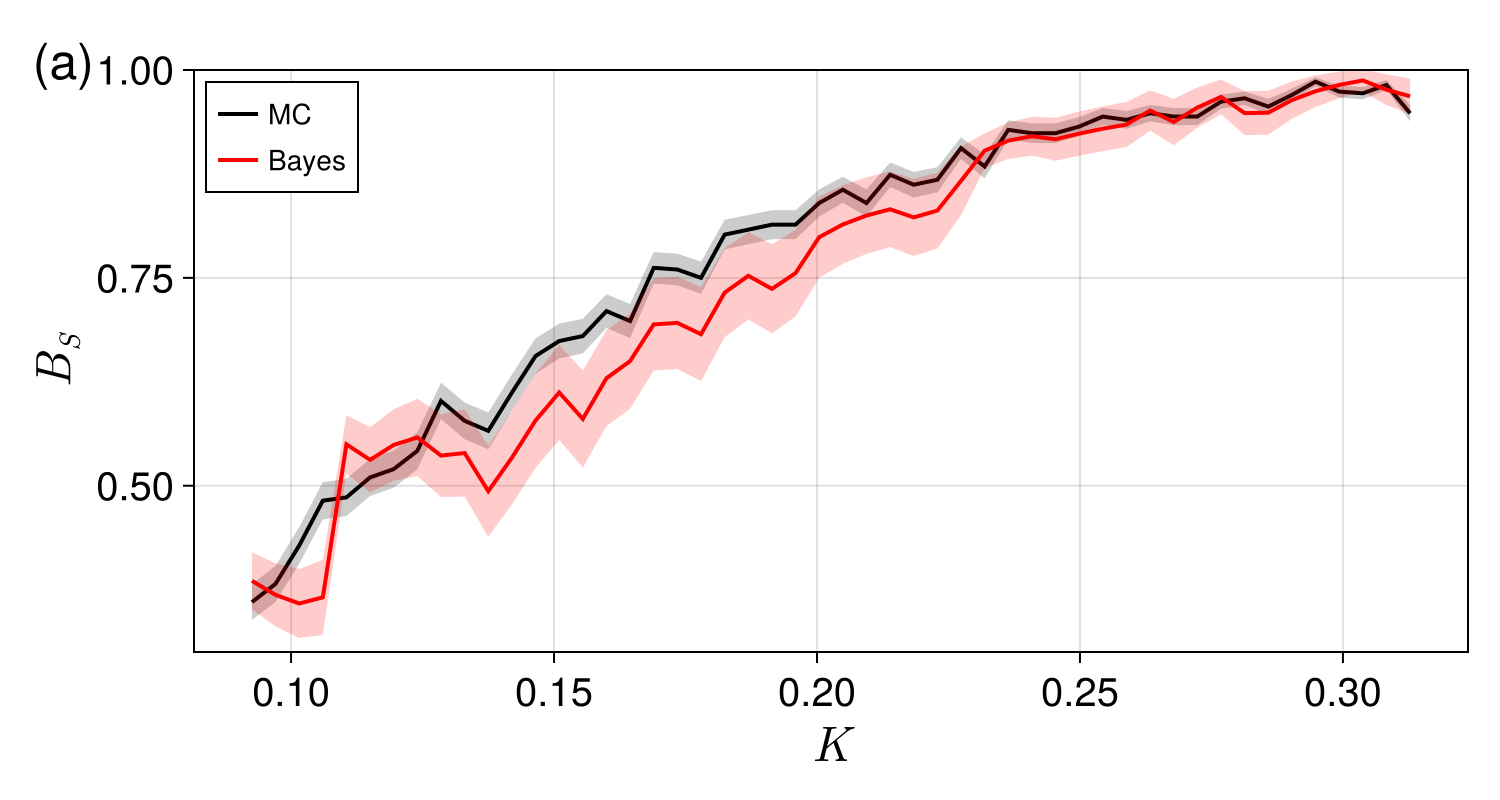}}
  \centerline{\includegraphics[width=0.50\textwidth]{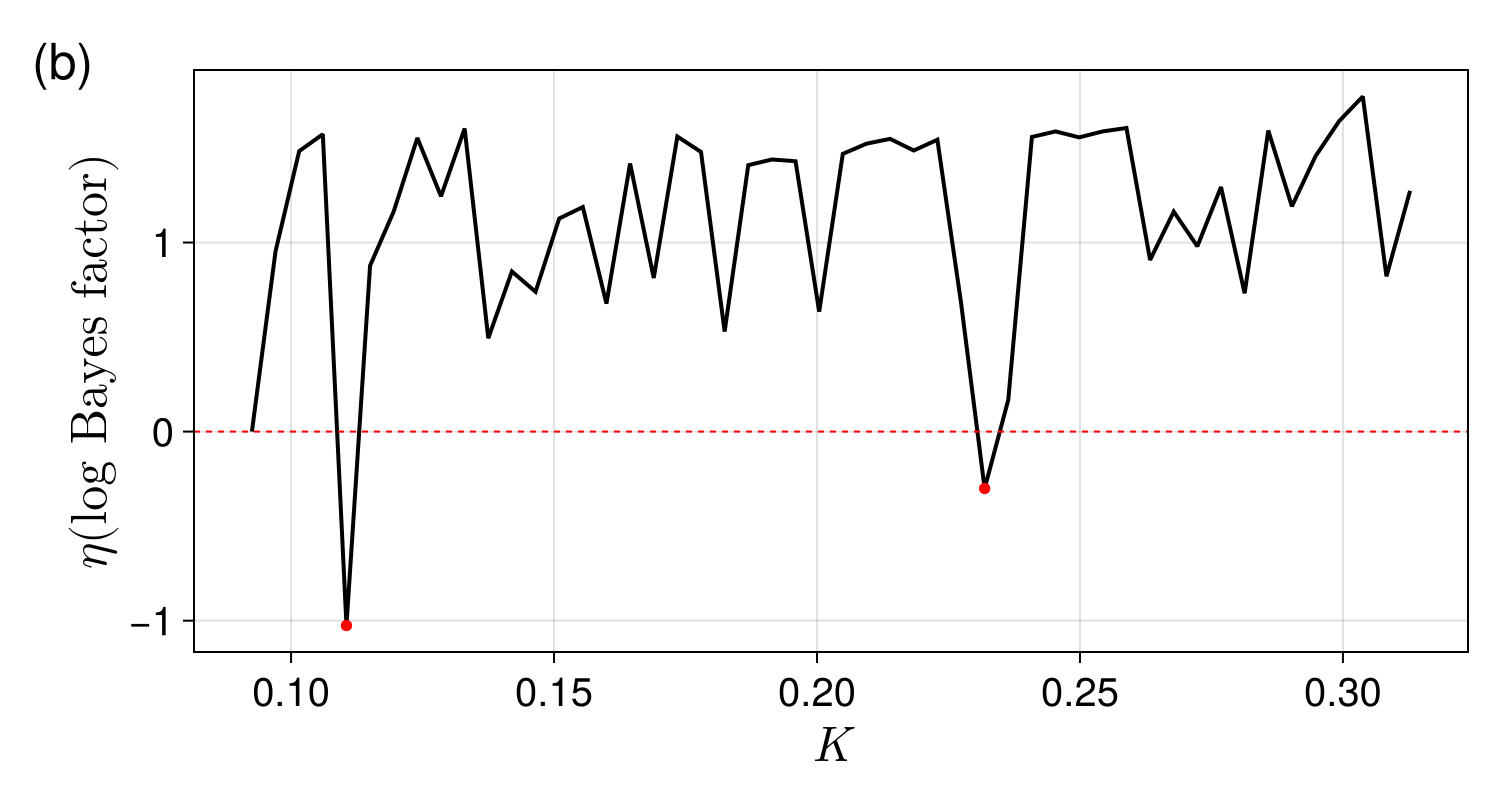}}
  \centerline{\includegraphics[width=0.50\textwidth]{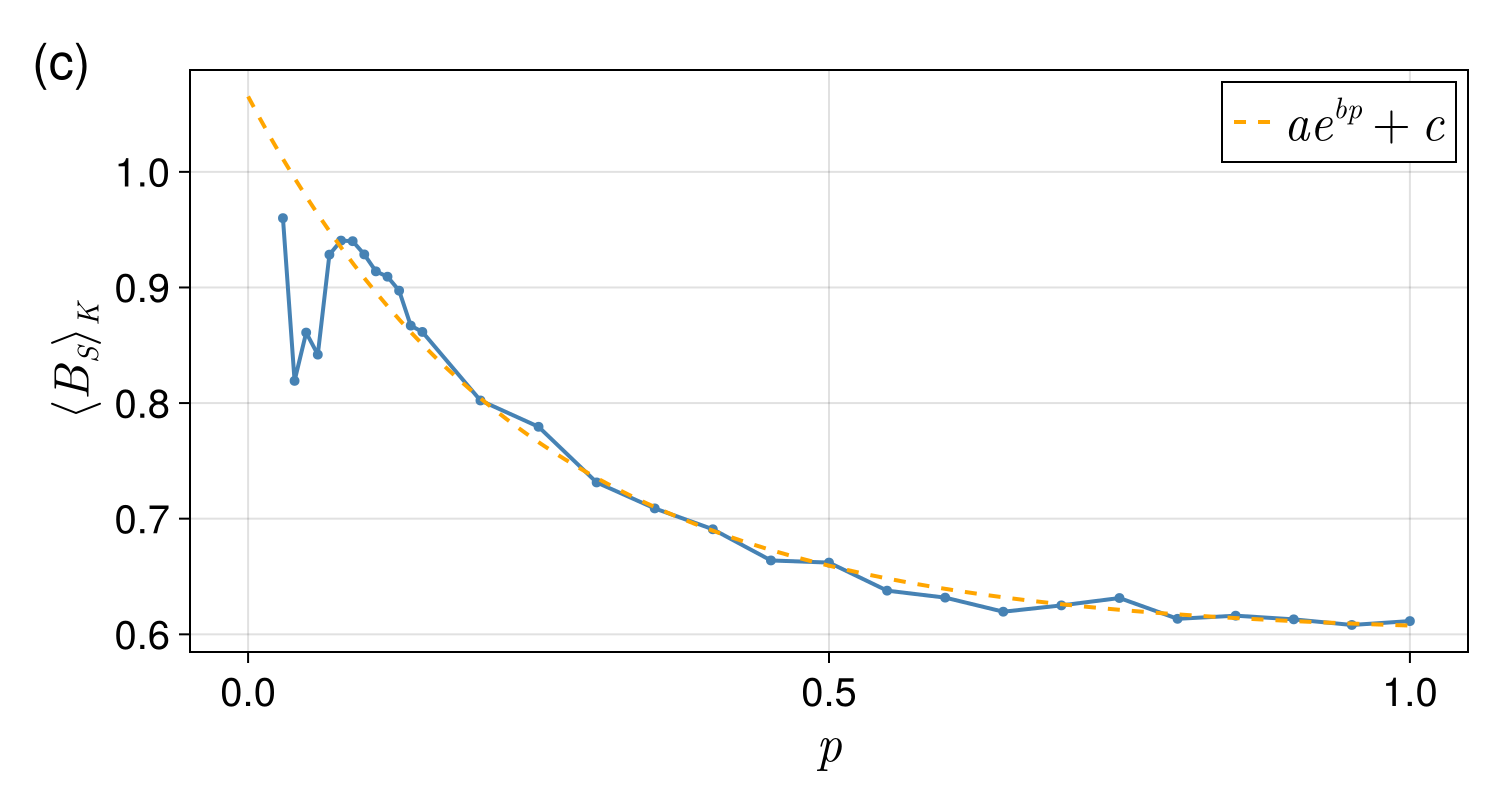}}
    \caption{Basin stability $B_S$ of the synchronous state for a network of $N=100$ coupled R\"ossler oscillators on a Watts-Strogatz graph ($\langle k\rangle =8$, $p=0.2$) as a function of coupling $K$ within the MSF stability window. The red line represents the Bayesian estimation, while the black line shows pure Monte Carlo simulations with 500 samples per coupling parameter $K$. The parameters of the BBT method were chosen as $N_s= 20$, $N_d =200$.  (a)~Fraction of initial conditions leading to synchronization $B_S$. (b)~Log Bayes factor $\eta$ monitoring the continuity hypothesis; red markers indicate autonomous reset. (c) Average of basin stability over the values of K as a function of the rewiring $p$. The simulations have been averaged over 10 different realizations of the network.}\label{fig:fig4}
\end{figure}

Figure~\ref{fig:fig4} (a) shows the results for a Watts-Strogatz network with rewiring probability $p=0.2$. As $K$ increases through the synchronization window, the fraction of initial conditions that synchronize (the basin stability $B_S$) grows from 0.3 to a plateau close to complete synchronization for all the initial conditions. The figure compares the dense estimation with $N_d = 500$ samples at each point (in black) with the Bayesian estimation in red. We can observe that the algorithm has detected the drift and triggered a reset at $K = 0.1$ and $K=0.2$. This is a good illustration of how the autonomous reset is triggered and recomputes the prior probabilities.  

In Fig.~\ref{fig:fig4} (b), the log Bayes factor $\eta$ remains positive throughout most of the sweep, although dense resampling is triggered on four occasions. For other values of $p$, the number of alarms varies between 1 and 7 for the simulations carried out. 

This application demonstrates two key aspects of the methodology. First, the Bayesian framework operates efficiently in a $300$-dimensional state space. Second, the continuation over the coupling strength $K$ provides a complete picture of the synchronization landscape for a fixed network topology.

For comparison with the original study, we provide the basin stability averaged over $K$ as a function of the wiring parameter $p$ in Fig.~\ref{fig:fig4} (c). For each wiring probability $p$, continuation was performed for 10 random network realizations for $N_s = 20$ and $N_d = 200$ samples. The simulation took about 8 hours on a standard laptop (Intel i7-1068 with 8 cores). The exponential decay is visible, but our results show a slightly different picture since the average basin volume of the synchronous state is higher. Since the original publication~\cite{menck_how_2013} does not explicitly discuss all details of the numerical integration procedure, we used a fifth-order stiffly stable Rosenbrock integrator with the full Jacobian matrix specified in order to avoid possible instabilities that we detected with other standard algorithms. The results indicate greater stability of the synchronous state than previously reported, although the qualitative conclusions remain unaltered.  

To complete the information on the performance of the algorithm, we have measured the root-mean-square difference between the synchronized-basin-volume estimates obtained with the Monte Carlo and Bayesian estimators for each generated network. The average over the wiring probability $p$ is $\langle d \rangle_p = 0.043$. 

Moreover the average alarm probability for the dense sampling process is $\bar{p}_a = 0.083$, leading to a computational gain $G = 5.75$ fairly constant across simulations. This average gain of a factor of $5.75$ demonstrates that the Bayesian framework achieves the accuracy of dense Monte Carlo integration at a fraction of the computational cost, which is particularly important in high-dimensional systems such as this one.

\section{Discussion and conclusion}
\label{sec:Concl}

The Bayesian Basin Tracking algorithm presented in this work successfully exploits the persistence of boundary structures in dynamical systems as a control parameter changes. By replacing deterministic grid evaluations with adaptive stochastic sampling, the method effectively circumvents the massive computational overhead associated with traditional basin continuation.

A primary strength of this framework is its system-agnostic nature. The algorithm relies entirely on the output of an independent oracle function $O(u)$ and a separate attractor-matching routine. Since evaluating this oracle ordinarily requires expensive numerical integration, isolating it as a "black box" allows BBT to be applied to discrete maps, continuous-time flows, and highly coupled networks alike.

Crucially, the method offers a robust pathway around the curse of dimensionality. While deterministic grid tessellations rapidly become computationally intractable as the dimension increases, our framework demonstrates that global basin metrics such as basin-volume fractions can be accurately estimated in massive state spaces (e.g., $d=300$) by treating the global bounding region as a single partition. The rigorous mathematical foundation of the Dirichlet-Multinomial model, combined with the closed-form variance of the Bayesian entropy, guarantees that the uncertainty of these estimates is continuously quantified.

The autonomous nature of the algorithm is governed by the Log-Bayes factor $\eta$, an information-theoretic sensor that actively monitors the validity of the continuity hypothesis. Rather than relying on arbitrary, user-defined thresholds, $\eta$ inherently balances the Occam advantage of historical memory against the Kullback-Leibler divergence of topological drift. 

A direct consequence of this data economy is a highly quantifiable computational gain over standard Monte Carlo continuation. At every parameter increment, a baseline brute-force scheme evaluates $N_d$ samples per partition. In contrast, our procedure draws only $N_s \ll N_d$ sparse samples, paying the dense cost $N_d$ exclusively on the fraction $\bar{p}_a$ of partitions flagged by the Log-Bayes factor. The resulting speed-up, $G \approx N_d / (N_s + \bar{p}_a N_d)$, operates near its theoretical upper limit $N_d/N_s$ in smooth regions of the parameter space and degrades gracefully toward unity only during severe boundary crises where dense resampling becomes mathematically necessary. Ultimately, this framework ensures that expensive computational resources are allocated precisely where the physics of the system demands them most.

\acknowledgments

This work was supported by grants from the Brazilian government agencies CNPq and CAPES. P. Haerter received partial financial support from CNPq (Grant No. 140920/2022-6), CAPES (Grant No. 88887.898818/2023-00) and FAPESP(Grant No. 2025/28656-4). M. A. F. Sanjuán acknowledges financial support from the Spanish State Research Agency (AEI) and the European Regional Development Fund (ERDF, EU) under Project No.~PID2023-148160NB-I00 (MCIN/AEI/10.13039/501100011033).


\appendix

\section{Marginal Likelihood of the Dirichlet-Multinomial Model}
\label{app:Ap1}

To evaluate the evidence for a given dynamical regime, we calculate the marginal likelihood of the observed counts $\mathbf{c} = \{c_1, \dots, c_\mathcal{N}\}$ given the Dirichlet hyperparameters $\boldsymbol{\alpha} = \{\alpha_1, \dots, \alpha_\mathcal{N}\}$. This requires marginalizing over the unknown probability simplex $\mathbf{p} = \{p_1, \dots, p_\mathcal{N}\}$ that describes the relative volumes of the basins of attraction within a specific phase space partition:
\begin{equation}
    P(\mathbf{c} \mid \boldsymbol{\alpha}) = \int_{\Delta^{\mathcal{N}-1}} P(\mathbf{c} \mid \mathbf{p}) P(\mathbf{p} \mid \boldsymbol{\alpha}) \, d\mathbf{p},
\end{equation}
where the integration is performed over the standard $(\mathcal{N}-1)$-simplex $\Delta^{\mathcal{N}-1} = \{ \mathbf{p} : \sum p_i = 1, p_i \ge 0 \}$. The likelihood of observing the count vector $\mathbf{c}$ given a fixed configuration $\mathbf{p}$ is governed by the multinomial distribution:
\begin{equation}
    P(\mathbf{c} \mid \mathbf{p}) = \frac{N_s!}{\prod_{i=1}^\mathcal{N} c_i!} \prod_{i=1}^\mathcal{N} p_i^{c_i},
\end{equation}
where $N_s = \sum c_i$ is the total number of sparse trajectories. Simultaneously, the prior density of the basin probabilities is given by the Dirichlet distribution:
\begin{equation}
    P(\mathbf{p} \mid \boldsymbol{\alpha}) = \frac{1}{\text{B}(\boldsymbol{\alpha})} \prod_{i=1}^\mathcal{N} p_i^{( \alpha_i - 1 )}.
\end{equation}
The normalization constant $\text{B}(\boldsymbol{\alpha})$ is the multivariate Beta function. Substituting these components into the integral, we obtain:
\begin{equation}
    P(\mathbf{c} \mid \boldsymbol{\alpha}) = \frac{N_s!}{\prod_{i=1}^\mathcal{N} c_i! \cdot \text{B}(\boldsymbol{\alpha})} \int_{\Delta^{\mathcal{N}-1}} \prod_{i=1}^\mathcal{N} p_i^{(c_i + \alpha_i) - 1} \, d\mathbf{p}.
\end{equation}
The integral over the simplex is precisely the definition of the multivariate Beta function for the updated parameters $\boldsymbol{\alpha}' = \mathbf{c} + \boldsymbol{\alpha}$. Consequently, the marginal likelihood simplifies to the ratio of Beta functions:
\begin{equation}
    P(\mathbf{c} \mid \boldsymbol{\alpha}) = \frac{N_s!}{\prod_{i=1}^\mathcal{N} c_i!} \frac{\text{B}(\boldsymbol{\alpha} + \mathbf{c})}{\text{B}(\boldsymbol{\alpha})}.
\end{equation}
When comparing the log-evidence of two distinct priors ($\boldsymbol{\alpha}^{(\text{prior})}$ vs. $\boldsymbol{\alpha}^{(\text{reset})}$) for the same observed counts $\mathbf{c}$, the multinomial coefficient is a constant factor that cancels out in the log Bayes factor calculation. Therefore, for the purposes of identifying a regime shift, we define the log-evidence $L(\boldsymbol{\alpha})$ in terms of the log beta function:
\begin{equation}
    L(\boldsymbol{\alpha}) = \ln \text{B}(\boldsymbol{\alpha} + \mathbf{c}) - \ln \text{B}(\boldsymbol{\alpha}).
\end{equation}
This formulation highlights that the evidence is simply the log-change in the normalization constant of the Dirichlet distribution after incorporating the sparse sample $\mathbf{c}$. For implementation, $\ln \text{B}(\boldsymbol{\alpha})$ is computed using the identity $\ln \text{B}(\boldsymbol{\alpha}) = \sum \ln \Gamma(\alpha_i) - \ln \Gamma(\sum \alpha_i)$ to maintain numerical stability across high-dimensional phase space boxes.

\section{Steady-state analysis of the Log-Bayes Factor}
\label{app:logBF}

This appendix derives an asymptotic approximation for $\mathbb{E}[\eta]$ using a first-order expansion for the history model ($\alpha_i \gg 1$) and Stirling's approximation for the reset model. Let $\mathbf{p} = (p_1,\dots,p_\mathcal{N})$ be the true probabilities and $\mathbf{c} \sim \mathrm{Multinomial}(N_s, \mathbf{p})$ the observed counts. The log-evidence is (see Appendix~\ref{app:Ap1}):

\begin{equation}
    L(\boldsymbol{\alpha}) = \sum_{i=1}^\mathcal{N} \bigl[\ln\Gamma(\alpha_i + c_i) - \ln\Gamma(\alpha_i)\bigr] - \bigl[\ln\Gamma(\alpha_0 + N_s) - \ln\Gamma(\alpha_0)\bigr].
    \label{eq:app_logevidence}
\end{equation}
The two competing models have the following parameters:
\begin{itemize}
    \item History: $\alpha_i^{(h)} = \rho N_s \hat{p}_i$,\quad $\alpha_0^{(h)} = \rho N_s$,\quad $\rho = \gamma/(1-\gamma)$.
    \item Reset: $\alpha_i^{(r)} = 1/2$,\quad $\alpha_0^{(r)} = \mathcal{N}/2$ \quad(Jeffreys prior).
\end{itemize}

\subsection{Lag between true and prior probabilities}

The next step is to establish the recurrence relation that the algorithm imposes on the Bayesian estimate probability $\hat p_i$. The relation (\ref{alpha_update}) on parameters $\alpha^{(n+1)}_i = \gamma \alpha^n_i + c_i$ can be transformed with the expectation Eq.~(\ref{exp_p_i}): $\alpha^n_i = \hat p^n_i \alpha_0$ and the counts are generated by the true probability $c_i = N_s p^n_i $. Given that the total pseudo count is $\alpha_0 = N_s/(1-\gamma)$ in the steady-state drift regime, we obtain the fading-memory recurrence $\hat{p}_i^{(n+1)} = \gamma\hat{p}_i^{(n)} + (1-\gamma)p_i(n)$ in the form of a linear filter. Unrolling and substituting a linear drift $p_i^{(n-k)} = p_i^{(n)} - k\delta\,\partial p_i/\partial a$ yields:
\begin{equation}
    \hat{p}_i^{(n)} = p_i^{(n)} - \rho\,\delta\,\frac{\partial p_i^{n}}{\partial a}.
    \label{eq:drift_lag_app}
\end{equation}

The difference between the true distribution and the prior is measured by the Kullback-Leibler divergence $D_{KL}$. Defining $\Delta_i = p_i - \hat p_i$, we expand the divergence to second order in $\Delta_i$~\cite{cover1999elements}:
\begin{equation}
    D_{KL}(\mathbf{p}\|\hat{\mathbf{p}}) \approx \dfrac{1}{2}\sum_{i=1}^\mathcal{N} \dfrac{\Delta_i^2}{\hat p_i}.
\end{equation}

Substituting the lag relation (\ref{eq:drift_lag_app}) into the KL divergence yields: 
\begin{equation}\label{eq:DKL_drift}
    D_{KL}(\mathbf{p}\|\hat{\mathbf{p}}) \approx \tfrac{1}{2}\rho^2\delta^2\sum_{i=1}^\mathcal{N} \frac{\left(\frac{\partial p_i}{\partial a}\right)^2}{\hat p_i}.
\end{equation}

\subsection{Estimation of $\mathbb{E}[L_{\text{hist}}]$ and $\mathbb{E}[L_{\text{reset}}]$}

In Eq.~(\ref{eq:app_logevidence}), each term can be studied as a difference of log-Gamma functions with integer arguments. We use the identity $\ln\Gamma(\alpha+c)-\ln\Gamma(\alpha) = \sum_{k=0}^{c-1} \ln(\alpha+k)$ and expand $\ln(\alpha+k) \approx \ln\alpha + k/\alpha$ since $\alpha \gg c$, to obtain
\begin{equation}
    \ln\Gamma(\alpha + c) - \ln\Gamma(\alpha) \approx c\ln\alpha + \frac{c(c-1)}{2\alpha}.
    \label{eq:exact_gamma_exp}
\end{equation}

Applying this approximation to both the category and normalization terms in Eq.~\eqref{eq:app_logevidence} with $\alpha_i^{(h)} = \rho N_s \hat{p}_i$, the $\ln(\rho N_s)$ factors cancel:
\begin{equation}
    L_{\text{hist}} \approx \sum_{i=1}^\mathcal{N} \left[ c_i \ln\hat{p}_i + \frac{c_i(c_i-1)}{2\rho N_s \hat{p}_i} \right] - \frac{N_s-1}{2\rho}.
\end{equation}
Taking the expectation using $\mathbb{E}[c_i] = N_s p_i$ and $\mathbb{E}[c_i(c_i-1)] = N_s(N_s-1)p_i^2$:
\begin{equation}
    \mathbb{E}[L_{\text{hist}}] \approx N_s \sum_{i} p_i \ln\hat{p}_i + \frac{N_s-1}{2\rho} \left( \sum_{i} \frac{p_i^2}{\hat{p}_i} - 1 \right).
\end{equation}
The first term equals $-N_s H(\mathbf{p}) - N_s D_{KL}(\mathbf{p}\|\hat{\mathbf{p}})$. The parenthesis is equal to $\sum_i \Delta_i^2/\hat p_i \approx 2D_{KL}(\mathbf{p}\|\hat{\mathbf{p}})$ according to the previous section. Combining:
\begin{equation}
    \mathbb{E}[L_{\text{hist}}] \approx -N_s H(\mathbf{p}) - N_s\,D_{KL}(\mathbf{p}\|\hat{\mathbf{p}}) \left( 1 - \frac{1}{\rho} \right).
    \label{eq:ELhist}
\end{equation}
where we have set $(N_s-1) \approx N_s$. The factor $1 - 1/\rho = (2\gamma-1)/\gamma$ modulates the drift penalty, which vanishes as $\gamma$ approaches $1/2$ from above (rapid forgetting, $\rho \to 1$). For the reset model $L_{\text{reset}}$ under a Jeffreys prior ($\alpha_i = 1/2$), Stirling's approximation applied to the category terms gives, for each $c_i \gg 1$:
\begin{equation}
    \ln\Gamma(c_i+\tfrac{1}{2}) - \ln\Gamma(\tfrac{1}{2}) \approx c_i\ln c_i - c_i + \tfrac{1}{2}\ln 2.
\end{equation}
Evaluating the expectation at the mean counts $\bar{c}_i = N_s p_i$ and applying Stirling to the normalization term $\ln\Gamma(\mathcal{N}/2+N_s)-\ln\Gamma(\mathcal{N}/2)$, the extensive $N_s\ln N_s$ and $N_s$ terms cancel, leaving
\begin{equation}
    \mathbb{E}[L_{\text{reset}}] \approx -N_s H(\mathbf{p}) - \frac{\mathcal{N}-1}{2}\ln\frac{2N_s}{\mathcal{N}}.
    \label{eq:ELreset}
\end{equation}

\subsection{Expectation $\mathbb{E}[\eta]$ in the normal drift regime}

Subtracting Eq.~\eqref{eq:ELreset} from Eq.~\eqref{eq:ELhist}, the entropy terms cancel:
\begin{equation}
    \mathbb{E}[\eta] \;\approx\; \underbrace{\frac{\mathcal{N}-1}{2}\ln\frac{2N_s}{\mathcal{N}}}_{\text{Occam advantage}} \;-\; \underbrace{N_s\,D_{KL}(\mathbf{p}\|\hat{\mathbf{p}})\!\left(1 - \frac{1}{\rho}\right)}_{\text{drift penalty}}. 
    \label{eq:eta_final}
\end{equation}
The Occam advantage reflects the structural economy of the history model with the scaling $\mathcal{N}\ln(N_s/\mathcal{N})$. The drift penalty is the cost of the tracking lag, attenuated by the factor $(1 - 1/\rho)$: for $\rho \gg 1$ (long memory) the full $D_{KL}$ penalty applies, while as $\rho \to 1$ the penalty vanishes because the prior becomes too weak to conflict with the data. Substituting the drift approximation for $D_{KL}$ from Eq.~\eqref{eq:DKL_drift}, the self-consistency condition $\mathbb{E}[\eta] > 0$ reduces to
\begin{equation}
    \frac{N_s\,\rho(\rho-1)\,\delta^2}{2}\,\sum_{i=1}^\mathcal{N} \frac{\left(\frac{\partial p_i}{\partial a}\right)^2}{\hat p_i}\;\ll\; \frac{\mathcal{N}-1}{2}\ln\frac{2N_s}{\mathcal{N}},
\end{equation}
where the factored form $\rho(\rho-1)$ makes the validity boundary $\rho > 1$ (i.e.\ $\gamma > 1/2$) immediately transparent. The expansion additionally requires $N_s p_i \gg 1$.

While Stirling's expansion mathematically assumes large sample sizes, the logarithmic scaling of the Occam advantage ensures this bound holds robustly even in the sparse sampling regime.

\end{document}